\documentclass[11pt,russian]{article}   
\usepackage{cmap}
\usepackage[T2A]{fontenc}
\usepackage[utf8]{inputenc} 
\usepackage[english]{babel}
\usepackage{amsthm}
\normalfont

\usepackage{amssymb,amsmath}
\usepackage{color,graphicx}

\makeatletter
\def\@seccntformat#1{\csname the#1\endcsname.\ } 
\makeatother

\newtheorem{theorem}{Theorem}
\newtheorem{conj}{Conjecture}

\begin{document}

\noindent
\textbf{\LARGE On perfect codes that do not include \\ Preparata-like codes%
\footnote{The work was funded by the Russian science foundation (grant 14-11-00555)}}
\date{}

\vspace*{5mm} \noindent
\textsc{D. S. Krotov, A. Yu. Vasil'eva} \hfill \texttt{krotov,vasilan@math.nsc.ru} \\
{\small Sobolev Institute of Mathematics, Novosibirsk, Russia} \\

\medskip

\begin{center}
\parbox{11,8cm}{\footnotesize
\textbf{Abstract.}
We show that for every length of form $4^k-1$,
there exists a binary $1$-perfect code that does not include any Preparata-like code.}
\end{center}

\baselineskip=0.9\normalbaselineskip

\section{Introduction}

The binary $1$-error-correcting perfect codes (below, perfect codes)
and the codes with the parameters of the Preparata codes
(below, Preparata codes)
are two remarkable infinite families of binary codes having many common properties.
Many of these properties are related to the fact that
any code with the parameters (length, size, code distance)
of the codes from the considered classes
induces a regular partition of the space.
Every Preparata code is included in some perfect code \cite{E:ZZS:73:Praparata}.

A perfect code can include several nonequivalent Preparata codes.
In the current correspondence, we consider the question
of the existence of perfect codes that does not include any Preparata code
(of course, we consider only lengths of form $4^n-1$,
for which Preparata codes exist \cite{E:Preparata:1968}).

As is known since \cite{E:Vas:nongroup_perfect},
the number of nonequivalent perfect codes grows double-exponentially with respect to the length
(a survey of some constructions can be found in \cite{E:Sol:2008:survey},
\cite{E:Hed:2008:survey}).
Before \cite{E:Vas:nongroup_perfect},
only the linear perfect code, the Hamming code, was known;
moreover because of some strict metrical invariant properties of the perfect codes
\cite{E:Lloyd}, \cite{E:ShSl}, it was conjectured that there are no other
$1$-error correcting perfect binary codes.

At the moment, it is known only two classes of Preparata codes.
The first class contains the original Preparata codes \cite{E:Preparata:1968}
and their generalizations
\cite{E:Dumer:1976}, \cite{E:BvLW}, \cite{E:vDamFDFla:2000}
(the last paper contains the most general known representation
of codes from this class).
These codes are nonlinear, but the including perfect code is linear, the Hamming code.
The second class is the class of $Z_4$-linear Preparata codes.
Strictly speaking, the corresponding
distance-$6$ extended Preparata codes are $Z_4$-linear.
The first known series of such codes was published in \cite{E:HammonsOth:Z4_linearity}.
As was shown in \cite{E:Kantor:2004symplecticsemifield},
the number of nonequivalent $Z_4$-linear extended Preparata codes of length
 $n=4^m$ grows over-polynomially in $n$ for almost all values of $m$.
All Preparata codes from the second class are includes in the same, up to equivalence,
perfect code, whose extension is $Z_4$-linear.
This perfect code is nonlinear if $n>15$ \cite{E:HammonsOth:Z4_linearity}.

So, for every length $n=4^m-1>15$, only two nonequivalent perfect codes
including Preparata codes are known.
The fraction of these codes among the set of all perfect codes is unknown.
It is naturally to conjecture that these fraction
is negligibly small; however,
before the current work,
there was no perfect code known that has length $4^m-1 \ge 63$
and guaranteedly does not include any
Preparata code.
In the current correspondence, we show a short proof of the existence of such perfect codes,
also hoping that this will attract an attention to the problem and stimulate
obtaining more essential results in this direction.

Note that for the length $15$ the situation is clear
(although it cannot be used to predict the asymptotic behavior).
The Preparata code of length $15$ is unique up to equivalence
and known as the Nordstrom-Robinson code, see, e.g., \cite{E:MWS};
this code is included in the Hamming code.
Hence, all nonlinear perfect codes of length $15$ do not include Preparata codes.

\section{Preliminaries}\label{Prelim}

We study codes in the $n$-\emph{dimensional binary Hamming space},
consisting from the set $Q_n$ of all binary $n$-tuples (words), with component-wise
modulo-$2$ addition and the Hamming metric.
The \emph{support} $\mathrm{supp}(\alpha)$ of the word $\alpha$ is the set of its nonzero positions;
the cardinality of the support of a word  $\alpha$ is its \emph{Hamming weight} $\mathrm{wt}(\alpha)$.
The  \emph{Hamming distance} $\rho(\alpha,\beta)$ between words $\alpha$ and $\beta$
is the Hamming weight of $\alpha+\beta$.
By $|\alpha|$, we denote the modulo-$2$ sum of the coordinates of the word $\alpha$.
For binary words $\alpha$ and $\beta$ (their lengths may be different),
$(\alpha,\beta)$ denotes their concatenation.

A set  $C\subseteq Q^n$ of $M$ words with mutual distance at least  $d$
is called a \emph{binary $(n,M,d)$ code}, i.e., a code of length $n$, size $M$, and distance $d$.
A code is called \emph{perfect (with distance $3$)} if the balls of radius $1$ centered
in the code words do not intersect and cover all $Q_n$.
It is straightforward from the definition that the minimal distance between codewords is $3$.
Perfect codes of length $n$ exist for every $n$ of form $n=2^t-1$
and do not exist for any other $n$ (because the cardinality of a radius-$1$ ball must divide the cardinality of $Q_n$).
For every $n=2^t-1$,
there exist a unique,
up to equivalence, linear
(i.e., closed with respect to addition) perfect code, the Hamming code.
In the half of the cases, namely, when $t$ is even,
there are Preparata codes of length $n=2^t-1$,
which are defined as the codes of distance $5$ and size $2^{n+1}/(n+1)^2$.
Every Preparata code is included in a unique perfect code~\cite{E:ZZS:73:Praparata}.

A \emph{Steiner triple system} of order $n$, or $\mathrm{STS}(n)$,
is a collection of $3$-subsets of $\{1,\ldots,n\}$, called \emph{blocks},
such that every pair of elements are included in exactly one block.
The cardinality of $\mathrm{STS}(n)$ is $n(n-1)/6$.
Given a perfect code $C$ and its codeword $\alpha$, 
the set of words $\beta$ of $C$ at distance $3$
from  $\alpha$ defines the Steiner triple system $T(\alpha)$
as follows:
$$T(\alpha) = \{\mathrm{supp}(\beta+\alpha) \ : \ \beta\in C, \ \rho(\alpha,\beta)=3\} .$$
If $C$ includes a Preparata code $P$ and $\alpha\in C\backslash P$, 
then this $\mathrm{STS}(n)$ has the subset
\begin{equation}\label{TPrep}
\{\mathrm{supp}(\beta+\alpha) \ : \ \beta\in P, \ \rho(\alpha,\beta)=3\},
\end{equation}
which partitions the set $\{1,\ldots,n\}$ into $3$-subsets \cite{E:ZZS:73:Praparata,SZZ:1971:UPC}.

A nonempty 
subset $R$ of a code $C$ if called an $i$-\emph{component} of $C$
if the code $(C\setminus R)\cup  (R+ \boldsymbol{e}^i)$,
where $\boldsymbol{e}^i$ is a word of weight $1$ with the $i$th nonzero coordinate, has the same parameters as $C$,
and no proper subset of $R$ satisfies this property.
Two words of a distance-$d$ code are called
$i$-\emph{close} if they differ in $d$ positions including $i$.
If an $i$-component $R$ contains a word $\alpha$, then $R$ contains all words $i$-close to $\alpha$
\cite{E:Sol88en}.
The following fact was proven in \cite{E:Tok:2004.en}.
If a perfect code $C$ includes a Preparata code $P$
and  $R$ is an $i$-component of $C$,
then
the set $P\cap R$ is a perfect code in the graph $(R,E)$,
where $E$ is the set of pairs of words from $R$ at distance $3$.

Let us describe the Vasil'ev construction \cite{E:Vas:nongroup_perfect}
of perfect codes.
Let $H_k$ be a Hamming code of length $k=2^t-1$,
and let $\lambda$ be an arbitrary function from $H_k$ to $\{0,1\}$;
then the set
$$ \{ ( \alpha,\alpha+\beta,|\alpha|+\lambda(\beta) ) \ : \ \alpha\in Q_k, \ \beta\in H_k\} $$
is a perfect code of length $n=2k+1$.
In the case $\lambda\equiv 0$, the resulting code is the Hamming code of length $n=2k+1$.

From this representation, we can see the following.
For every word $(\alpha,\alpha+\beta,|\alpha| )$ from the Hamming code,
the $n$-close words are the words 
$(\alpha,\alpha+\beta,|\alpha|+1)+\boldsymbol{e}^j+\boldsymbol{e}^{k+j}$, $j=1,\ldots,k$.
Hence, for every word $\beta$ from $H_k$, the set
\begin{equation}\label{R-comp}
R=R(\beta) = \{ ( \alpha,\alpha+\beta,|\alpha| ) \ : \ \alpha\in Q_k\}
\end{equation}
is an $n$-component (called \emph{linear}).
The set $\{R(\beta) \,:\, \beta \in H_k\}$ of all such components forms a partition of the Hamming code of length $n$.
Each of these $n$-components can be independently replaced by its switching
$R'= R+\boldsymbol{e}^n$;
as a result we can obtain a large number of new perfect codes.

\section{The main result}\label{MainRes}

We are going to show that some Vasil'ev codes,
namely, the codes obtained from the Hamming code
by switching one component,
do not include any Preparata codes.
\begin{theorem} \label{nPr}
Let $n=4^t-1$ and $\beta\in H_{(n-1)/2}$.
Then, no Preparata codes are included in the perfect code
\begin{equation}\label{C}
C = C(\beta)= (H_n\setminus R)\cup R' ,
\end{equation}
where the component $R=R(\beta)$ is defined in (\ref{R-comp})
and $R'=R+ \boldsymbol{e}^n$ is its switching.
\end{theorem}

{\bf Proof}.
Assume that there exists a Preparata code $P\subseteq C$.
Let a word $\boldsymbol{x}$ of $P$ belongs to the ``switched'' component; i.e.,
$\boldsymbol{x}\in P\cap R'$.
Such a word exists because a Preparata code has a nonempty intersection
with each $i$-component of the including perfect code
(see, for ex., \cite{E:Tok:2004.en}; this also follows from the uniqueness of a perfect code including $P$).
At distance $3$ from $\boldsymbol{x}$, we choose a codeword
$\boldsymbol{y}$ of $C$ that
does not belong to the switched component,
$\boldsymbol{y}\in C\setminus R'$.
Such $\boldsymbol{y}$ exists because there are $n(n-1)/6$
codewords at distance $3$
from $\boldsymbol{x}$, but only $(n-1)/2$ of them
are in the same $n$-component as $\boldsymbol{x}$.
Note that $\boldsymbol{y} \not\in P$.

Without loss of generality we can assume that
$\boldsymbol{y} = \boldsymbol{0}=(0,\ldots,0)$
(if it is not the case, we consider $C+\boldsymbol{y}$ instead of $C$ and $P+\boldsymbol{y}$ instead of $P$).
Then $\mathrm{wt}(\boldsymbol{x})=3$;
let $\{i,j,k\}$ be the support of $\boldsymbol{x}$.
For $R'=R(\beta)+ \boldsymbol{e}^n$ to contain
$\boldsymbol{x}$,
the condition
$\mathrm{wt}(\beta)=3$ should be satisfied (see (\ref{R-comp})).
In this case, it is easy to see from (\ref{R-comp}) that $R'$ contains only $4$ words of weight $3$, with the supports
\begin{equation}\label{STS_C}
\{i,j,k\}, \ \{i\,j'\,k'\}, \ \{i',j,k'\}, \ \{i'\,j'\,k\},
\end{equation}
 for some different
 $i'$, $j'$, $k'$ from $\{1\ldots,n\}\setminus\{i,j,k\}$.
The triple $\{i,j,k\}$
belongs to the partition $T$
of $\{1,\ldots,n\}$
into triples corresponding to the weight-$3$ words of the Preparata code $P$,
see (\ref{TPrep});
the other three triples from
(\ref{STS_C}), obviously, do not belong to $T$.

Let us consider the Steiner triple system $S_{H_n}$
corresponding to the weight-$3$ words of the Hamming code $H_n$.
Since $$H_n=(C\setminus R')\cup R,$$
we see that the Steiner triple system $S_{H_n}$ is obtained from the Steiner triple system $S_C$ of $C$ by replacing the triples (\ref{STS_C}),
corresponding to words of  $R'$, by the triples
$$\{i,j,k'\}, \ \{i,j',k\}, \ \{i',j,k\}, \ \{i',j',k'\},$$
corresponding to words of  $R'$.
So, we have $T\setminus\{\{i,j,k\}\}\subseteq S_{H_n}$,
and the words with the supports
from $T\setminus\{\{i,j,k\}\}$ belong to the Hamming code.
As the Hamming code is linear, we have
$$\boldsymbol{u} = \sum_{\mathrm{supp}(\boldsymbol{z})\in T\setminus \{\{i,j,k\}\}} \boldsymbol{z} \in H_n .$$
Since a perfect code is antipodal and
$\boldsymbol{y} = \boldsymbol{0}=(0,\ldots,0)\in H_n$,
we also have $\boldsymbol{1}=(1,\ldots,1)\in H_n$;
hence,
$\boldsymbol{v}=\boldsymbol{1}+\boldsymbol{u}\in H_n$.
However, from the definition of $\boldsymbol{u}$ we get
$\mathrm{supp}(\boldsymbol{v})= \{i,j,k\}$ and, consequently,
$\boldsymbol{v}=\boldsymbol{x}\notin H_n$.
A contradiction.

\section{Conclusion}

We have shown that some Vasil'ev codes do not include Preparata codes.
Taking into account a local character of the proof,
we can conclude that the same is true for a large number of nonequivalent
Vasil'ev codes.
This way does not yield directly that almost all
 Vasil'ev codes do not include Preparata codes;
 however, it is natural to conjecture this.
 We finish with two conjectures based on common properties of the two perfect codes
 that are known to include Preparata codes.

 \begin{conj}
  All the codewords of weight $3$ ($4$) of a (extended) perfect code including a Preparata code form a linear Steiner triple (respectively, quadruple) system,
  i.e., a system equivalent to that is formed by the weight-$3$ (weight-$3$) words of the linear Hamming code.
 \end{conj}
 \begin{conj}
  All $i$-components (for every $i$) of a perfect code including a Preparata code are linear, i.e., equivalent to the set $\{(x,x,|x|)\}$.
 \end{conj}


\providecommand\href[2]{#2} \providecommand\url[1]{\href{#1}{#1}}
  \def\DOI#1{{\small {DOI}:
  \href{http://dx.doi.org/#1}{#1}}}\def\DOIURL#1#2{{\small{DOI}:
  \href{http://dx.doi.org/#2}{#1}}}

\end{document}


\noindent
\textbf{\LARGE О совершенных кодах,\\ не включающих кодов Препарата%
\footnote{Исследование выполнено за счет гранта Российского научного фонда (проект № 14-11-00555)}}
\date{}

\vspace*{5mm} \noindent
\textsc{Д. С. Кротов, А. Ю. Васильева} \hfill \texttt{krotov,vasilan@math.nsc.ru} \\
{\small Институт математики им. С.Л. Соболева СО РАН,  \\
 Новосибирск, Россия} \\

\medskip

\begin{center}
\parbox{11,8cm}{\footnotesize
Для любой длины $4^k-1$ указан совершенный код, не содержащий ни одного кода с параметрами кода Препарата. }
\end{center}

\baselineskip=0.9\normalbaselineskip

\section{Введение}

Двоичные совершенные коды, исправляющие одну ошибку (далее --- совершенные коды)
и коды с параметрами кода Препарата (далее --- коды Препарата)
--- два исключительных бесконечных семейства двоичных кодов,
имеющих много общих свойств.
Многие из этих свойств связаны с тем, что любой код с параметрами (длина, мощность, кодовое расстояние) из рассматриваемых классов порождает регулярное разбиение
пространства.
Всякий код Препарата включается в некоторый, притом единственный, совершенный код \cite{ZZS:73:Praparata}.

Совершенный код может включать несколько неэквивалентных
кодов Препарата. В настоящей заметке рассмотрен вопрос
существования совершенных кодов, не включающих коды Препарата
(разумеется, рассматриваются только коды длин вида $4^n-1$, для других
коды Препарата не существуют \cite{Preparata:1968}).

Как известно со времен работы  Васильева \cite{Vas:nongroup_perfect.ru},
число неэквивалентных совершенных кодов растет дважды экспоненциально
по отношению к длине
(обзор некоторых других конструкций см. в \cite{Sol:2008:survey},
\cite{Hed:2008:survey}).
До той работы был известен только линейный совершенный код Хэмминга,
и, ввиду весьма жестких метрических инвариантов,
которым должны удовлетворять совершенные коды \cite{Lloyd}, \cite{ShSl},
высказывались предположения,
что других совершенных кодов, исправляющих одну ошибку, нет.

В настоящий момент известно два класса кодов с параметрами кодов Препарата.
Первый класс включает оригинальные коды Препарата
\cite{Preparata:1968} и их обобщения
\cite{Dumer:1976.ru}, \cite{BvLW}, \cite{vDamFDFla:2000}
(последняя работа содержит наиболее общее из известных представлений кодов этой серии).
Сами эти коды нелинейны,
но включающий их совершенный код линейный --- код Хэмминга.
Второй класс --- $Z_4$-линейные коды Препарата.
Точнее, $Z_4$-линейными являются расширенные коды Препарата с расстоянием $6$.
Первая серия таких кодов была опубликована в работе \cite{HammonsOth:Z4_linearity}.
В статье \cite{Kantor:2004symplecticsemifield}
показано, что число неэквивалентных
$Z_4$-линейных расширенных кодов Препарата
длины $n=4^m-1$ растет сверхполиномиально по $n$
для почти всех значений $m$.
Все коды Препарата из второго класса содержатся в одном и том же, с точностью до эквивалентности, совершенном коде, расширение которого является $Z_4$-линейным. Этот совершенный код нелинеен при $n>15$ \cite{HammonsOth:Z4_linearity}.

Таким образом, для каждой длины $n=4^m-1>15$ известно только два неэквивалентных совершенных кода, включающих коды Препарата.
Доля же таких кодов среди множества всех совершенных кодов неизвестна. Хотя и кажется естественным предположить, что эта доля
ничтожно мала, до настоящей работы не было известно ни одного совершенного кода длины $4^m-1 \ge 63$, про который известно, что он не включает код Препарата.
В данной заметке мы приводим короткое доказательство существования таких кодов, надеясь также, что это привлечет внимание к проблеме и стимулирует получение более существенных результатов в данном направлении.

Заметим, что для длины $15$ ситуация ясна,
хотя из нее и нельзя делать выводы об асимптотическом поведении:
код Препарата этой длины единственный с точностью до эквивалентности
и известен как код Нордстрома-Робинсона, см. напр. \cite{MWS.ru}. Этот код включается в линейный совершенный код. Поэтому все нелинейные совершенные коды этой длины не включают коды Препарата.

\section{Предварительные сведения}\label{Prelim}

Мы изучаем коды в $n$-\emph{мерном двоичном пространстве Хэмминга}, состоящем из множества $Q_n$
всех двоичных наборов длины $n$ с покомпонентным
сложением по модулю $2$ и определенной на них метрикой Хэмминга.
\emph{Носителем} $\mathrm{supp}(\alpha)$ вершины $\alpha$ называется множество ее ненулевых позиций, мощность носителя -- это \emph{вес Хэмминга} $\mathrm{wt}(\alpha)$ вершины $\alpha$;
тогда \emph{расстояние Хэмминга} $\rho(\alpha,\beta)$ между двумя вершинами $\alpha$ и $\beta$  -- это вес Хэмминга их суммы $\alpha+\beta$.
Через $|\alpha|$ будем обозначать сумму по модулю $2$ координат набора $\alpha$.
Если $\alpha$ и $\beta$ -- двоичные наборы, то через $(\alpha,\beta)$ будем обозначать их конкатенацию.

Множество $C\subseteq Q^n$, состоящее из $M$ элементов, попарные расстояния между которыми не меньше $d$, называется \emph{двоичным кодом с параметрами} $(n,M,d)$, т.е длины $n$, мощности $M$, с расстоянием $d$. Код называется \emph{совершенным (с расстоянием $3$)}, если шары радиуса $1$ с центрами
в вершинах кода не пересекаются и покрывают весь куб. Действительно, из определения вытекает,
что минимальное растояние кода равно $3$. Совершенные коды длины $n$ существуют
для всех $n$ вида $n=2^t-1$, и только для них.
Каждый раз есть единственный с точностью до изоморфизма
линейный (т.е. являющийся линейным пространством) совершенный код -- код Хэмминга.
В половине случаев существования совершенных кодов,
а именно, для длины $n$ вида $n=2^t-1$,
когда $t$ четно, существуют также \emph{коды Препарата}, определяемые как
имеющие кодовое расстояние $5$ и максимально большую мощность,
равную $2^{n+1}/(n+1)^2$.
Любой код Препарата  содержится в некотором,
притом единственном, совершенном коде~\cite{ZZS:73:Praparata}.

\emph{Cистемой троек Штейнера порядка} $n$ ($STS(n)$), называется такая система троек
элементов множества $\{1,\ldots,n\}$, что любая пара элементов встречается в точности в одной тройке; мощность $STS(n)$ равна $n(n-1)/6$.
Множество вершин $\beta$ произвольного совершенного кода, расположенных на расстоянии $3$ от произвольной фиксированной вершины $\alpha$ этого кода, определяет систему троек Штейнера $T(\alpha)$, где:
$$T(\alpha) = \{\mathrm{supp}(\beta+\alpha) \ : \ \beta\in C, \ \rho(\alpha,\beta)=3\} .$$
Если данный совершенный код $C$ содержит в качестве подмножества
некоторый код Препарата $P$ и $\alpha\in C\setminus P$, то
эта система имеет подмножество
\begin{equation}\label{TPrep}
\{\mathrm{supp}(\beta+\alpha) \ : \ \beta\in P, \ \rho(\alpha,\beta)=3\},
\end{equation}
являющееся разбиением множества $\{1,\ldots,n\}$ на тройки \cite{ZZS:73:Praparata,SZZ:1971:UPC.ru}.

Непустое подмножество $R$ кода $C$ называется его $i$-\emph{компонентой},
если множество $(C\setminus R)\cup  (R+ \boldsymbol{e}^i)$,
где $\boldsymbol{e}^i$ -- единичный вектор с ненулевой $i$-й координатой,
является кодом с теми же параметрами,
причем никакое собственное подмножество из $R$ этому свойству не удовлетворяет.
Две вершины кода с расстоянием $d$ называются
$i$-\emph{близкими}, если они находятся на расстоянии $d$ и отличаются в $i$-й координате.
Подмножество кода является его $i$-компонентой, если вместе с любой своей вершиной содержит
и все $i$-близкие к ней \cite{Sol88}.
Отметим следующий полезный факт \cite{Tok:2004.ru}.
Если $C$ -- совершенный код, содержащий код Препарата $P$, и $R$ -- произвольная $i$-компонента
кода $C$, то в графе $(R,E)$, где $E$ -- множество пар вершин из $R$ на расстоянии $3$,
множество $P\cap R$ является совершенным кодом.

Опишем конструкцию Васильева \cite{Vas:nongroup_perfect.ru} нелинейных совершенных кодов.
Пусть $k=2^t-1$, \ $H_k$ -- код Хэмминга длины $k$ и
пусть $\lambda: H_k \rightarrow \{0;1\}$ -- произвольная функция. Тогда множество
$$ \{ ( \alpha,\alpha+\beta,|\alpha|+\lambda(\beta) ) \ : \ \alpha\in Q_k, \ \beta\in H_k\} $$
является совершенным кодом длины $n=2k+1$. В случае $\lambda\equiv 0$ этот код является кодом Хэмминга длины $n=2k+1$.

Из этого представления вытекает,
что для произвольной вершины вида $(\alpha,\alpha+\beta,|\alpha| )$ из кода Хэмминга
$n$-близкими к ней являются все вершины вида
$(\alpha,\alpha+\beta,|\alpha|+1)+\boldsymbol{e}^j+\boldsymbol{e}^{k+j}$, $j=1,\ldots,k$, и только они.
Поэтому для каждой вершины $\beta\in H_k$ из кода Хэмминга длины $k$ множество
\begin{equation}\label{R-comp}
R=R(\beta) = \{ ( \alpha,\alpha+\beta,|\alpha| ) \ : \ \alpha\in Q_k\}
\end{equation}
является $n$-компонентой (компоненты такого вида будем называть \emph{линейными}),
причем их совокупность $\{R(\beta) \,:\, \beta \in H_k\}$ образует разбиение кода Хэмминга длины $n$.
Каждую из этих компонент можно независимо заменить в коде на ее сдвиг $R'= R+\boldsymbol{e}^n$ и таким образом получить большое число новых совершенных кодов.

\section{Основной результат}\label{MainRes}

Покажем, что некоторые из кодов Васильева
не могут содержать ни одного кода Препарата, а именно,
совершенные коды, отличающиеся от кода Хэмминга сдвигом одной компоненты.
\begin{Rtheorem} \label{nPr}
Пусть $n=4^t-1$ и $\beta\in H_{(n-1)/2}$.
Тогда не существует кода Препарата, вложимого в совершенный код
\begin{equation}\label{C}
C = C(\beta)= (H_n\setminus R)\cup R' ,
\end{equation}
где компонента $R=R(\beta)$  определена в \rm{(\ref{R-comp})}
и $R'=R+ \boldsymbol{e}^n$ -- ее сдвиг.
\end{Rtheorem}

{\bf Доказательство}.
Предположим, что существует некоторый код Препарата $P\subseteq C$.
Пусть вершина $\boldsymbol{x}$ кода Препарата лежит в <<сдвинутой>> компоненте:
$\boldsymbol{x}\in P\cap R'$. Такая вершина существует, поскольку код Препараты имеет непустое пересечение с каждой $i$-компонентой объемлющего совершенного кода
(см., например, \cite{Tok:2004.ru}; кроме того, это следует из единственности совершенного кода, включающего $P$).
На расстоянии $3$ от $\boldsymbol{x}$ выберем вершину $\boldsymbol{y}$ совершенного кода $C$,
не принадлежащую ни коду Препарата, ни <<сдвинутой>> компоненте:
$\boldsymbol{y}\in (C\setminus P)\setminus R'$.
Такая вершина $\boldsymbol{y}$ существует,
поскольку на расстоянии 3 от данной вершины $\boldsymbol{x}$ совершенного кода
находится в точности  $n(n-1)/6$ вершин этого кода и лишь $(n-1)/2$ из них
лежат в той же компоненте, что и вершина $\boldsymbol{x}$.

Без ограничения общности можно считать, что $\boldsymbol{y} = \boldsymbol{0}=(0,\ldots,0)$.
(Если это не так, будем рассматривать $C+\boldsymbol{y}$ вместо $C$ и $P+\boldsymbol{y}$ вместо $P$.)
Тогда $\mathrm{wt}(\boldsymbol{x})=3$, и пусть носитель $\mathrm{supp}(\boldsymbol{x})$ -- это тройка $\{i,j,k\}$.
Чтобы компонента $R'=R(\beta)+ \boldsymbol{e}^n$ могла содержать вершину $\boldsymbol{x}$
веса $3$, должно быть выполнено условие
$\mathrm{wt}(\beta)=3$ (см. (\ref{R-comp})).
В этом случае из (\ref{R-comp}) легко увидеть,
что $R'$ содержит всего четыре вершины веса $3$, их носители имеют
следующий вид (для некоторых различных $i',j',k'\in \{1,\ldots,n\}\setminus\{i,j,k\}$):
\begin{equation}\label{STS_C}
\{i,j,k\}, \ \{i,j',k'\}, \ \{i',j,k'\}, \ \{i',j',k\} .
\end{equation}
Кроме того, тройка $\{i,j,k\}$ входит в некоторое разбиение $T$
 множества $\{1,\ldots,n\}$ на тройки, отвечающие вершинам веса $3$ из кода Препарата $P$ (см. (\ref{TPrep})),
остальные три тройки из (\ref{STS_C}) в него, очевидно, не входят.

Рассмотрим систему троек Штейнера $S_{H_n}$, соответствующую вершинам веса $3$
кода Хэмминга $H_n$, из которого был построен код $C=(H_n\setminus R)\cup R'$.
Поскольку $$H_n=(C\setminus R')\cup R,$$
то $S_{H_n}$ получается из системы троек Штейнера $S_C$ кода $C$ заменой троек (\ref{STS_C}),
определяемых вершинами из $R'$, на тройки, определяемые вершинами из $R$:
$$\{i,j,k'\}, \ \{i,j',k\}, \ \{i',j,k\}, \ \{i',j',k'\} .$$
Таким образом, $T\setminus\{\{i,j,k\}\}\subseteq S_{H_n}$ и вершины
с носителями из $T\setminus\{\{i,j,k\}\}$ лежат в коде Хэмминга.
В силу линейности кода Хэмминга
$$\boldsymbol{u} = \sum_{\mathrm{supp}(\boldsymbol{z})\in T\setminus \{\{i,j,k\}\}} \boldsymbol{z} \in H_n .$$
Поскольку совершенный код антиподален и
$\boldsymbol{y} = \boldsymbol{0}=(0,\ldots,0)\in H_n$,
то и $\boldsymbol{1}=(1,\ldots,1)\in H_n$, а потому
и $\boldsymbol{v}=\boldsymbol{1}+\boldsymbol{u}\in H_n$.
Но из определения вершины $\boldsymbol{u}$ следует, что
$\mathrm{supp}(\boldsymbol{v})= \{i,j,k\}$ и поэтому $\boldsymbol{v}=\boldsymbol{x}\notin H_n$.
Противоречие.

\section{Заключение}

Мы показали, что некоторые коды Васильева не включают ни одного кода Препарата. Учитывая локальный характер доказательства, можно сделать вывод, что то же верно для большого числа неэкваивалентных кодов Васильева.
Хотя этим способом и не удается напрямую доказать, что почти все коды Васильева не включают коды Препарата, это также естественно предположить.
В заключение сформулируем две гипотезы на основе общих свойств двух совершенных кодов, про которые известно, что они включают коды Препарата.

Гипотеза 1. Вершины веса $3$ ($4$) любого (расширенного) совершенного кода, включающего код Препарата, образуют линейную систему троек (соответственно, систему четверок) Штейнера, т.е. эквивалентную содержащейся в коде Хэмминга.

Гипотеза 2. Все $i$-компоненты (для любого $i$) совершенного кода, включающего код Препарата, линейны, то есть эквивалентны множеству $\{(x,x,|x|)\}$.
